\documentclass[12pt]{article}
\usepackage{epsf,epsfig,amsmath}
\begin{document}

\title{Exclusion of Time in Mermin's Proof of Bell-Type Inequalities}

\author{Karl Hess$^1$ and Walter Philipp$^2$}

\date{$^1$ Beckman Institute, Department of Electrical Engineering
and Department of Physics,University of Illinois, Urbana, Il 61801
\\ $^{2}$ Beckman Institute, Department of Statistics and Department of
Mathematics, University of Illinois,Urbana, Il 61801 \\ }
%\date{\today}
\maketitle

\begin{abstract}

Mermin states that his nontechnical version of Bell's theorem
stands and is not invalidated by time and setting dependent
instrument parameters as claimed in one of our previous papers. We
identify deviations from well-established protocol in probability
theory as well as mathematical contradictions in Mermin's argument
and show that Mermin's conclusions are therefore not valid: his
proof does not go forward if certain possible time dependencies
are taken into account.

\end{abstract}

\section{Introduction}

We have presented a critique \cite{hp} of a nontechnical version
of Bell's theorem presented by Mermin \cite{mermin} and forthwith
referred to as MI. This critique was based on the introduction of
hidden time and setting dependent instrument parameters in
addition to the parameters considered by Bell and others
\cite{bellbook}. Mermin has responded to this critique
\cite{recmer}, \cite{eurom} and has attempted to show that our
time and setting dependent instrument parameters fail to undermine
his reasoning and that our extended parameter space ``collapses"
onto his. We demonstrate below that Mermin \cite{eurom} (referred
to as MII) has not properly considered the role of time and the
possible stochastic independence of the family of our instrument
parameters from the source parameter. When these and other factors
are taken into account it becomes clear that our extended
parameter space does indeed render his proof invalid.

Owing to the complexity of the problem, we will revert to notation
similar to that defined in our previous publications \cite{hp},
\cite{hpp1}-\cite{hpp3}. However, we denote the random variables
with capital letters and use the lower case for the values that
these variables may assume. To facilitate the discussion, we
provide a one to one correspondence of Mermin's notation and ours,
at least as far as possible. We consider random variables $A=\pm1$
in station $S_1$ and $B=\pm1$ in station $S_2$ that describe spin
measurements and are indexed by instrument settings that are
characterized by three-dimensional unit vectors ${\bf a}, {\bf b},
{\bf c}$ in both stations. MI introduces no precise counterpart
for the random variables $A, B$ and only considers green and red
detector flashes. These green and red flashes correspond then to
values that $A, B$ assume: $+1$ which we can identify with green
and $-1$ which we can identify with red. The key assumption of MI
in his original proof \cite{mermin} and of Bell and others
\cite{bellbook} is that the random variables $A, B$ depend only on
the setting in the respective station and on another random
variable $\Lambda$ that characterizes the particles emitted from a
common source. Instead of $\Lambda$, MI uses ``instruction sets"
e.g. GGR meaning flash green for settings ${\bf a}, {\bf b}$
(which MI actually labels $1, 2$) and flash red for setting ${\bf
c}$ (labelled $3$ by MI). In our notation this means, for
instance, that for a certain value $\lambda$ (of the parameter
$\Lambda$) that corresponds to the instruction set GGR we have
$A_{\bf a}( \lambda) = A_{\bf b}( \lambda) = +1$ and $A_{\bf c}(
\lambda) = -1$. The possible choices of $\Lambda$ are restricted
by MI \cite{mermin} and Bell \cite{bellbook} invoking Einstein
locality: The source parameter $\Lambda$ (Mermin's original
\cite{mermin} instruction sets, as well as their frequency of
occurrence) does not depend on the settings.

\section{Mermin's Proof and our Critique}

Mermin's main argument is based on the following facts
\cite{eurom}. ``The data accumulated in many runs of this
experiment have two important features:

\begin{itemize}

\item[i)] In those runs in which the detectors happen to have been
given the same settings, the lights always flash the same color.

\item[ii)] If all runs are examined without reference to the
settings of the detectors, the pattern of flashes is completely
random, in particular, the colors flashed are equally likely to be
the same or different."

\end{itemize}

We first analyze Mermin's original \cite{mermin} proof, but using
our notation instead of his. The following Table \ref{TA:ma}
summarizes the eight possible instruction sets and the nine
possible different $AB$ products which are used in MI, MII as a
model for EPR-experiments.
\begin{table}[ht]
  \begin{center}
    \begin{tabular}{|l||r|r|r||r|r|r||r|r|r|}\hline
      ${\Lambda}$ &
${A_{\bf a}}{B_{\bf a}}$ & ${A_{\bf a}}{B_{\bf b}}$ & ${A_{\bf
a}}{B_{\bf c}}$ & ${A_{\bf b}}{B_{\bf a}}$ & ${A_{\bf b}}{B_{\bf
b}}$ & ${A_{\bf b}}{B_{\bf c}}$ & ${A_{\bf c}}{B_{\bf a}}$ &
${A_{\bf c}}{B_{\bf b}}$ & ${A_{\bf c}}{B_{\bf c}}$ \\ \hline
      RRR & $+1$ & $+1$ & $+1$ & $+1$ & $+1$ & $+1$ & $+1$ & $+1$ & $+1$\\ \hline
      RRG & $+1$ & $+1$ & $-1$ & $+1$ & $+1$ & $-1$ & $-1$ & $-1$ & $+1$\\ \hline
      RGR & $+1$ & $-1$ & $+1$ & $-1$ & $+1$ & $-1$ & $+1$ & $-1$ & $+1$\\ \hline
      GRR & $+1$ & $-1$ & $-1$ & $-1$ & $+1$ & $+1$ & $-1$ & $+1$ & $+1$\\ \hline
      GGR & $+1$ & $+1$ & $-1$ & $+1$ & $+1$ & $-1$ & $-1$ & $-1$ & $+1$\\ \hline
      GRG & $+1$ & $-1$ & $+1$ & $-1$ & $+1$ & $-1$ & $+1$ & $-1$ & $+1$\\ \hline
      RGG & $+1$ & $-1$ & $-1$ & $-1$ & $+1$ & $+1$ & $-1$ & $+1$ & $+1$\\ \hline
      GGG & $+1$ & $+1$ & $+1$ & $+1$ & $+1$ & $+1$ & $+1$ & $+1$ & $+1$\\ \hline
   \end{tabular}
   \caption{Possible (but exclussive) $AB$ products}\label{TA:ma}
  \end{center}
\end{table}
According to point i) listed above, the columns of ${A_{\bf
a}}{B_{\bf a}}$, ${A_{\bf b}}{B_{\bf b}}$ and ${A_{\bf c}}{B_{\bf
c}}$ have all entries $+1$. The main point of MI, MII is then,
that each of the 8 rows of the Table \ref{TA:ma} contains at least
five entries $+1$ and at most 4 entries $-1$. This conclusion is
obtained by considering each entry in the remaining six columns
separately. As stated in MII p 144 ``Since each of the nine
possible pairs of settings is equally likely in any run, any of
the six instruction sets in which both colors appear (for example,
GGR) will result in the same color flashing $5/9$ of the time".
This amounts to averaging each of the eight rows of $\pm 1$ of
Table \ref{TA:ma}. Using our notation we can complete this step of
Mermin's argument in a slightly more transparent way. Consider the
row sums in Table \ref{TA:ma} for each single one of the eight
possible instruction sets (i.e. for each given value $\lambda_s,
s=1,...,8$ that $\Lambda$ may assume). This sum equals
\begin{equation}
{\sum_{row}}(\lambda_s):= A_{\bf a}B_{\bf a} + A_{\bf a}B_{\bf b}
+ A_{\bf a}B_{\bf c} +A_{\bf b}B_{\bf a} +A_{\bf b}B_{\bf b} +
A_{\bf b}B_{\bf c} + A_{\bf c}B_{\bf a} + A_{\bf c}B_{\bf b} +
A_{\bf c}B_{\bf c} \label{eum1}
\end{equation}
or
\begin{equation}
{\sum_{row}}(\lambda_s) = (A_{\bf a} + A_{\bf b} + A_{\bf
c})\cdot(B_{\bf a} + B_{\bf b} + B_{\bf c}) = (A_{\bf a} + {A_{\bf
b} + A_{\bf c})}^2 \label{eum2}
\end{equation}
because from i) given above it follows that $A_{\bf i} = B_{\bf
i}$ for ${\bf i} = {\bf a}, {\bf b}, {\bf c}$. Each pair of
settings occurs with probability $1/9$. Thus one obtains for the
average of each row
\begin{equation}
{\frac {1} {9}}{\sum_{row}}(\lambda_s) \geq {\frac {1} {9}}
\label{eum3}
\end{equation}
as the square equals either $1$ or $9$. This, according to MII,
``is incompatible with feature ii) of the data".

On the surface, all of this appears to be quite trivially correct.
There are, however, subtle contradictions in MII. It is stated on
p 143 (point 2)of \cite{eurom} that we are dealing with
``conjectured relations among various hypothetical outcomes of a
single experiment to be performed at a single time in one of
several different versions (which is what Bell's theorem is
about)" and not with ``outcomes of several different versions of
an experiment, all of which were actually performed at various
different times". According to this statement, there are no {\it
data} to be considered because Bell's theorem is only about
``conjectured relations..." and therefore the statement that i) is
incompatible with feature ii) of the {\it data} makes no sense.

This contradiction in Mermin's writings has its origins in the
fact that Table \ref{TA:ma} can not be directly linked to the
experimental data of any EPR experiment because the construction
and use of the table does not follow well established protocol of
probability theory. Feller \cite{feller} states ``If we want to
speak about experiments or observations in a theoretical way and
without ambiguity, we first must agree on the simple events
representing the thinkable outcomes; {\it they define the
idealized experiment}...... By definition {\it every
indecomposable result of the (idealized) experiment is represented
by one, and only one, sample point}. The aggregate of all sample
points will be called the {\it sample space}." Following Feller's
prescription for the case of EPR experiments of the type described
above, we define the simple (or indecomposable) events that form
the sample space $\Omega$ to be the 72 triplets of the form
$(\lambda_s; {\bf i}, {\bf j})$ where $\lambda_s, s = 1,...,8$ is
one of the eight instruction sets and $\bf i$, $\bf j$ can assume
the values $\bf a$, $\bf b$, or $\bf c$. Substituting these 72
triplets into the functions $A$ and $B$ and subsequently forming
the products $A.B$ we obtain the entries in Table \ref{TA:ma}.

The all important question is: {\it Is it possible to sample Table
\ref{TA:ma}}? Reformulated in mathematical terms this question
becomes: Is ${A_{\bf I}}(\Lambda).{B_{\bf J}}(\Lambda)$ a random
variable and what is its expectation value?

One certainly can sample the column sums of Table \ref{TA:ma}. For
example one could instruct the experimenters in both stations to
keep their choices of settings fixed throughout the duration of
the experiment. In the special case that each instruction set
$\lambda_s$ occurs with equal probability $\frac {1} {8}$ we
obtain nine expressions of the form ${\frac {1}
{8}}{\sum_{s=1}^8}{A_{\bf i}}({\lambda_s}).{B_{\bf
j}}({\lambda_s})$. This possibility is a special case of the
statistical argument discussed in section 3 below. However, it is
impossible to sample the row sums of Table \ref{TA:ma}, which of
course is the linchpin in Mermin's argument given above. Indeed it
is impossible to perform experiments or take observations for nine
different randomly chosen pairs of settings and to simultaneously
have the guarantee that the $\lambda_s$ remain all the same. Thus
\begin{equation}
Y := {\frac {1} {8}}{\sum_{{\bf i},{\bf j}={\bf a},{\bf b},{\bf
c}}}A_{\bf i}(\Lambda).B_{\bf j}(\Lambda) \label{eumc2.1}
\end{equation}
is not a random variable. To see this recall that the instruction
set $\Lambda$ is a random variable defined on the sample space
$\Omega$, that is $\Lambda = \Lambda(\omega)$ where $\omega \in
\Omega$ signifies any of the 72 simple (indecomposable) events
\cite{feller}. However, at a given time only one of the nine
products $A.B$ listed in Table \ref{TA:ma} can be measured.
Measurement of each of the nine products will require nine
separate experiments. Thus there will be nine different $\omega$'s
governing these nine experiments. Hence there is no guarantee that
the values of $\Lambda$ at these nine $\omega$'s will be the same,
nor is there a guarantee that the six terms $A_{\bf a}(\Lambda),
A_{\bf b}(\Lambda),..., B_{\bf c}(\Lambda)$ will be the same at
each occurrence. Thus there is no guarantee that $Y$ as given by
Eq.(\ref{eumc2.1}) will be a function of a single $\omega,$ only.
Consequently $Y$ is not a random variable. Hence, an application
of Fubini's theorem on double integration shows that $A_{\bf
I}(\Lambda).B_{\bf J}(\Lambda)$ is not a random variable and Table
\ref{TA:ma} cannot be sampled.

These considerations show, that the mathematical operations in MI,
MII that link Table \ref{TA:ma}, Eq.(\ref{eum3}) and the
experimental data do not properly include the subtle distinctions
that are necessary to define random variables. If one wants to
apply the inequality of Eq.(\ref{eum3}) to the statistics of the
experimental data one needs to make sure that all assumptions that
constitute the physical and mathematical model are (i) consistent
with well established mathematical protocol, (ii) consistent with
the procedures of the physical experiment, (iii) are general
enough to describe the physical experiment and, most importantly,
(iv) lead to a sampling of the complete Table \ref{TA:ma} by the
experimental procedure. It is the purpose of this paper to show
that the assumptions in MII do not fulfill these four requirements
and that the inclusion of setting and time dependent instrument
parameters prevents the proof that Table \ref{TA:ma} is
necessarily sampled by EPR experiments. These factors together
show that Eq.(\ref{eum3}) is not applicable to the statistics of
EPR experiments.

We first show that the parameter space used by Mermin is not
general enough to describe EPR experiments. MII claims on p 145
\cite{eurom} that explicit consideration of time can be excluded
``because it is obviously irrelevant". We interpret this statement
to mean that Eqs.(\ref{eum1})-(\ref{eum3}) can be proven and the
sampling of Table \ref{TA:ma} can be established with and without
consideration of time and setting dependent instrument parameters.
Therefore, we first ask the general question whether time
dependencies can somehow be absorbed into the instruction sets of
MII. Bell supporters often maintain that in each of their proofs
$\Lambda$ is totally arbitrary. In this context it is frequently
claimed that time is certainly ``known" also at the source and
therefore one must be able to combine any value $\lambda$ that
$\Lambda$ assumes with the time of measurement $t_{meas}$ to form
a new source parameter $\bar{\Lambda}$ with values $\bar{\lambda}
= (\lambda, t_{meas})$. However, such a combination runs into the
following serious mathematical contradiction. In all Bell-type
proofs, at a certain particular step of the proof, one must have
the same $\lambda$ for a sequence of different setting pairs. For
example, in Mermin's proof as given in Eqs.(\ref{eum1}) -
(\ref{eum3}) above, one must have the same $\lambda_s$ for all
nine pairs of settings ($({\bf i}, {\bf j}) = {\bf a}, {\bf b},
{\bf c}$). Yet, the measurement times $t_{{\bf i}{\bf j}}$ for
different pairs of settings must all be different by a strictly
positive amount. Thus, $\lambda$ and $t_{meas}$ follow different
and contradicting mathematical requirements and can not be
combined into a new parameter $\bar{\Lambda}$. Thus, the procedure
used in MI, MII to prove Bell-type inequalities prevents the
general use of time and time dependencies and therefore restricts
the parameter space. The setting and time dependence of our
instrument parameters represents an additional generalization of
the parameter space and introduces further unsurmountable problems
to Bell-type proofs. Note that we have indexed the time of
measurement by the settings on both sides. This fact indicates
only that the given time of measurement $t_{{\bf i}{\bf j}}$ has
been determined by the choice of the given settings ${\bf i},{\bf
j}$ just before the measurement occurred. It certainly does not
mean that any nonlocal dependence on both settings is introduced
either into $A$ or into $B$. The circumstance, however, that
determines how $t_{{\bf i}{\bf j}}$ is actually chosen shows that
this choice does depend on both settings and illustrates once more
the special role of time for EPR experiments.

We now subdivide our proof (that the assumptions leading to Table
\ref{TA:ma} are mathematically and physically inconsistent with
the EPR experiments or not general enough to describe them) into
two parts (a1) and (b1). The first part (a1) deals with arguments
that are made for the rows of Table \ref{TA:ma} and the second
(b1) for the columns.

\begin{itemize}

\item[(a1)] In actual experiments it is impossible to take measurements
for nine different pairs of settings and simultaneously to have
the guarantee that the $\lambda_s$ are all the same. Therefore,
the relevance of Table \ref{TA:ma} to statistical considerations
that apply also to the real experiments must first be proven. Only
then can a contradiction to the statistical properties of the data
follow from Eq.(\ref{eum3}).

Indeed, and this is just a variant of Mermin's characterization of
Bell's theorem as quoted above, it is often argued that
Eq.(\ref{eum3}) must be valid because one could have made a
measurement at the same time with a different setting chosen by a
person with free will, all in spite of the fact that only one term
of Eq.(\ref{eum1}) can be measured at a given time. We have no
problem with the counterfactual argument that we can consider what
would have happened if for any given term of a row any of the
other eight pairs of settings were in force at the analyzer
stations. But it is plainly inadmissible, indeed
``countersyntaxial" \cite{cst}, to sum, without further
justification, all nine possible outcomes for a given instruction
set. As an illustration consider the following example. Suppose
that a restaurant serves three main dishes $A_{\bf a}$, $A_{\bf
b}$ and $A_{\bf c}$ as well as three side dishes $B_{\bf a}$,
$B_{\bf b}$ and $B_{\bf c}$ for a total of nine items on the menu.
Assign the value $+1$ to the combination $[A_{\bf i}, B_{\bf j}]$
for ${\bf i}, {\bf j} = {\bf a}, {\bf b}, {\bf c}$ if it causes no
ill effect for a given patron $\lambda_s$ and assign the value
$-1$ to the combination $[A_{\bf i}, B_{\bf j}]$ if it does. We
can not imagine any scenario where it would make sense to add the
results of the nine possible outcomes. Yet this is exactly the
reasoning behind the statement in MII that Bell's theorem is
dealing with ``conjectured relations". What is the justification?
For this, MII seems to point to statistics, as did Bell before. We
agree that indeed the argument can be saved by statistical
reasoning. However, this salvage operation is only possible for a
very limited set of parameters and not for our extended parameter
space that properly includes time.

\item[(b1)] One can attempt to avoid the row argument and just
argue with the columns of Table \ref{TA:ma}. In order to do this
one needs to assume that each column arises from the same setting
independent random variable $\Lambda$. In certain special cases
such an argument based on the assumptions of MII will work, for
instance, if each of the eight instruction sets are assumed with
probability $\frac {1} {8}$. However, such an argument clearly
does not work if each column would be subject to a different
$\Lambda$ (instruction set random variable). If setting dependent
instrument parameter random variables are involved in the
formation of the instructions then there is no reason why, for
example, the joint probability density of these parameter random
variables should be the same for each of the nine different pairs
of settings. Nor can these joint densities be factorized because
of the possible time dependencies on both sides. We will show in
section 4 how Einstein local different (for different settings)
joint distributions can be constructed by use of two independent
computers with equal clock time.

\end{itemize}

The above discussion indicates various degrees of failure in the
reasoning of MII (and actually in all standard Bell-type proofs).
Counterfactual arguments, as for example that one could have
chosen another setting, may not yet render the proofs invalid.
Countersyntaxial arguments, such as the addition of various
outcomes that can not co-exist simultaneously, are wrong from the
point of view of logic. Conclusions resulting from such arguments
may still be correct, but certainly are suspicious. Proofs based
on self-contradictory arguments, such as the combination of time
and source parameter, always are fatal. The proof in MI, taking
into account the source parameter only, is in the best of
circumstances just counterfactual and, for a limited class of
parameters, can be made rigorous by using the statistical argument
and {\it reordering} (section 3).

The MII proof, attempted with setting and time dependent
instrument parameters, contains the whole gamut of problems
ranging from counterfactual to countersyntaxial to containing
mathematical contradictions. To show the contradictions in the MII
proof in the clearest possible way, we proceed below as follows.
We do not start with Eqs.(\ref{eum1})-(\ref{eum3}) or Table
\ref{TA:ma} but instead with the statistical arguments that permit
the use of Eqs.(\ref{eum1})-(\ref{eum3}) or Table \ref{TA:ma}. We
show then that our extended parameter space does not collapse onto
Mermin's and that the statistical arguments can not be made for
our extended parameter space.

\section{The Statistical Argument and Reordering}

We reformulate Bell's main statistical idea for Mermin's original
example \cite{mermin}. Assume that the number $M$ of values that
the hidden parameter $\Lambda$ can assume is finite (extension to
a countably infinite number is easy). In MI, MII we have $M = 8$.
Denote this set of values by $\{\lambda_s\}$ and the probability
$P(\Lambda = \lambda_s)$ by $p_s$, all with $s = 1,...,M$. Let N
be the number of experiments performed. If $N$ is large, then we
have by the strong law of large numbers that with probability $1$
the number of occurrences of $\lambda_s$ is approximately equal to
$N\cdot{p_s}$, $s = 1,...,M$. Assume now with MII that each pair
of settings $({\bf a}, {\bf a})$, $({\bf a}, {\bf b})$, etc.
occurs with probability $1/9$. The fact that the measurements can
only be made in sequence plays no role, because now one can {\it
reorder} the measurements into rows for a given $\lambda_s$ and
one could recognize such rows in the experiments if $\lambda_s$
could somehow be made visible. To be more specific, for each $s =
1,...,M$ we would have observed about ${\frac {1} {9}}Np_s$ times
the value $\lambda_s$ with each of the nine pairs of settings
because of the stochastic independence of $\Lambda$ and the
settings. Thus, after {\it reordering} the data we would have
accumulated about ${\frac {1} {9}}Np_s$ rows of the form as given
in Eq.(\ref{eum1}), each of them corresponding to the value
$\lambda_s$. For each of these rows Eq.(\ref{eum3}) holds. Of
course, there may be some terms left over. However, by the strong
law of large numbers, the number of such incomplete rows is
negligible for large $N$. This possibility of {\it reordering} is
equivalent to the point of view that the column sums in Table
\ref{TA:ma} and in a metaphorical sense Table \ref{TA:ma} itself
are actually sampled or accumulated by the measurement process. We
recall in passing that tables similar to Table \ref{TA:ma} and
their rows are used in various proofs of the Bell inequalities,
often without regard to the main points of Bell which show under
which circumstance and how these rows can be used to prove these
inequalities. To just claim that because Bell's inequalities
follow from Table \ref{TA:ma} that this will prove Bell's
inequalities is circular logic. The proof of Bell's theorem is
only then complete if a proof is given that in some appropriate
way Table \ref{TA:ma} is actually statistically sampled by the
EPR-type experiments.

\section{Time and Setting Dependent Instrument Variables}

Key to all Bell-type proofs is that the source parameter $\Lambda$
is independent of the instrument settings in the stations which is
guaranteed by the design of the experiments (delayed choice of
settings after particles have left the source). We have shown that
the argument involving Eqs.(\ref{eum1}) and (\ref{eum3}) is
countersyntaxial, except when salvaged by a statistical argument
that involves reordering of the elements of these equations. As a
consequence, this type of proof of Bell's theorem eliminates only
a small class of hidden variables. Bell and his followers maintain
that in their proofs they eliminate all Einstein local parameters.
We show below that the proof of the statistical argument given in
section 3 comes to a halt when Einstein local time and setting
dependent instrument parameters are included even in the special
case where the distribution of the source parameter $\Lambda$ does
not depend on time.

We introduce these setting and time dependent instrument variables
by using the example of two independent computers $C_1$ in $S_1$
and $C_2$ in $S_2$ with identical clock time. The computers
contain evaluation routines $A, B$ that map the source parameter
$\Lambda$ and local instrument parameters $\Lambda_{\bf a}^*(t)$
(that are generated by computer $C_1$) and $\Lambda_{\bf
b}^{**}(t)$ (that are generated by computer $C_2$) into $\pm 1$.
These instrument parameters can be thought of as arbitrarily
complicated numerical routines that supply output $\pm 1$ from the
input of setting and time, the only condition being that for equal
time and setting we must have, using Mermin's convention,
\begin{equation}
A_{\bf i} = B_{\bf i} \text{   for   } {\bf i} = {\bf a}, {\bf b},
{\bf c} \label{eum4}
\end{equation}
We maintain that any proof of Bell-type theorems needs to be able
to accommodate such parameters if it should be taken seriously.
Notice that we have now at least seven random variables $\Lambda$,
$\Lambda_{\bf i}^*(t)$ and $\Lambda_{\bf j}^{**}(t)$ with $({\bf
i}, {\bf j}) = {\bf a}, {\bf b}, {\bf c}$; in fact we have
$\Lambda$ plus six families of random variables.

Mermin \cite{eurom} claims that our model collapses onto his
because of Eq.(\ref{eum4}). We shall show presently that this
claim is false. Let $t_{{\bf i}{\bf j}}^{(l)}$ be the times  when
$A$ and $B$ are measured with settings ${\bf i} = {\bf a}, {\bf
b}$, or ${\bf c}$ in $S_1$ and ${\bf j} = {\bf a}, {\bf b}$, or
${\bf c}$ in $S_2$. Here $l = 1,2,...,L$. By the strong law of
large numbers we can assume that $L$ is about the same for all
nine pairs of settings $({\bf i}, {\bf j})$. Eq.(\ref{eum4}) thus
becomes:
\begin{equation}
A_{\bf i}( \Lambda, {\Lambda_{\bf i}^{*}}(t_{{\bf i}{\bf
i}}^{(l)})) = B_{\bf i}( \Lambda, {\Lambda_{\bf i}^{**}}(t_{{\bf
i}{\bf i}}^{(l)}))\label{eum5}
\end{equation}
On page 145 of MII it is stated about microsettings (our
instrument parameters) that ``no matter how strongly correlated
the microsettings may otherwise be, the expanded instruction sets
for that run must assign the same color to every microsetting
underlying a given setting....i.e. for each of the three
settings". Expressed mathematically this statement results in:
\begin{equation}
A_{\bf a}( \Lambda) = A_{\bf a}( \Lambda, {\Lambda_{\bf
a}^{*}}(t_{{\bf a}{\bf a}}^{(l)})) = A_{\bf a}( \Lambda,
{\Lambda_{\bf a}^{*}}(t_{{\bf a}{\bf b}}^{(l)})) = A_{\bf a}(
\Lambda, {\Lambda_{\bf a}^{*}}(t_{{\bf a}{\bf
c}}^{(l)}))\label{eum6}
\end{equation}
as well as
\begin{equation}
B_{\bf a}( \Lambda) = B_{\bf a}( \Lambda, {\Lambda_{\bf
a}^{**}}(t_{{\bf a}{\bf a}}^{(l)})) = B_{\bf a}( \Lambda,
{\Lambda_{\bf a}^{**}}(t_{{\bf b}{\bf a}}^{(l)})) = B_{\bf a}(
\Lambda, {\Lambda_{\bf a}^{**}}(t_{{\bf c}{\bf a}}^{(l)}))
\label{eum6a}
\end{equation}
and similar relations with cyclical exchange of ${\bf a}, {\bf b},
{\bf c}$. Of course Eqs.(\ref{eum5}), (\ref{eum6}) and
(\ref{eum6a}) imply
\begin{equation}
A_{\bf a}( \Lambda) = B_{\bf a}( \Lambda)\label{eum6b}
\end{equation}
Thus MII postulates that if two functions $A$ and $B$ assume equal
values on a finite set of points $t_{{\bf i}{\bf i}}^{(l)}, l =
1,...,L$ then $A$ and $B$ must be identical constants (remember
that the measurement times must all be different for different
setting pairs). We are not aware of any theorem of calculus that
would allow for such a sweeping statement. Nor does the physics
involved permit any such conclusions. The instrument parameters
are locally determined and will depend on the settings. Einstein
locality and the delayed choice of the settings does not impose
any further conditions in contrast to the situation with a source
parameter only (that must not depend on the settings).

Mermin argues for the extended parameter space exactly in the same
way he argues when only a source parameter is used. He states
about the parameters of the extended space on page 145
\cite{eurom}: ``these must exist in every run, whether or not the
detectors do end up with the same setting....and the particles
have to be prepared for every one of these possibilities when they
leave the source". The whole point of our work is, of course, that
we include {\it setting} and time dependent instrument parameters
in the extended instruction set. Therefore the statement ``these
must exist...whether or not the detectors do end up with the same
setting" is lacking logic. Mermin's statement would only make
sense for the source parameter $\Lambda$, because the instrument
settings are rapidly changed and are chosen only after the
particles have left the source. Therefore, it would be correct to
state for source parameters that ``these must exist...whether or
not the detectors do end up with the same settings".

For instrument parameters (detector parameters), the flapping
about of the instrument settings is, of course, locally known and
{\it they are determined} by the settings at which the detectors
do end up. It is important to realize that the random variables
$A_{\bf i}( \Lambda, \Lambda_{{\bf i}t}^{*})$ and $B_{\bf j}(
\Lambda, \Lambda_{{\bf j}t}^{**})$ can {\it not} be rewritten as
${\bar{A}}_{\bf i}(\bar{\Lambda})$ and ${\bar{B}}_{\bf
j}(\bar{\Lambda})$ with the time $t$ being absorbed into
$\bar{\Lambda}$. We have demonstrated in section 2 that this
argumentation is not valid even when only source parameters and
time are considered. A fortiori, this argument fails when time and
setting dependent instrument parameters are added. Therefore the
statement of MII that ``The expanded instruction sets of Hess and
Philipp must thus collapse back to the instruction sets of my
example" is refuted.

We now turn to the question whether our expanded instruction sets,
assuredly different from those of MII and Bell, will permit the
all important {\it statistical argument}.

\begin{itemize}

\item[(a2)] The reason why the statistical argument as the one
given in section 3 (justifying the use of the same $\lambda$ in
the rows) is invalid if time and setting dependent instrument
parameters are included is the following. Assume just for
simplicity the special case that the distribution of the source
parameter $\Lambda$ does not depend on time. Then there will be
still about ${\frac {1} {9}}Np_s$ rows of data ${A_{\bf i}}{B_{\bf
j}}$ $({\bf i},{\bf j} = {\bf a}, {\bf b}, {\bf c})$ each
corresponding to the values $\lambda_s$ that the source parameter
$\Lambda$ can assume. However, the terms
\begin{equation}
A_{\bf i}( {\lambda_s}, {\lambda_{\bf i}^{*}}(t_{{\bf i}{\bf
j}}^{(u)}))B_{\bf j}( {\lambda_s}, {\lambda_{\bf j}^{**}}(t_{{\bf
i}{\bf j}}^{(u)}))\label{eum7}
\end{equation}
with $u = 1,2,...,{\frac {1} {9}}Np_s$ may be all different
because they contain different measurement times for different
settings. Here, the difference of the role of time and the role of
the source parameter becomes very clear. While the measurement
times can under no circumstance be equal for different settings,
the information $\lambda_s$ that is sent out from the source and
relates to the spin must be independent of the settings and
appears ${\frac {1} {9}}Np_s$ times. As mentioned before MII
maintains (``time is irrelevant") that the pair $\lambda, t$ can
be concatenated to new source parameter values $\bar{\lambda}$.
However, the mathematical conditions for the values $\lambda_s$
and the measurement times $t_{{\bf i}{\bf j}}$ contradict each
other and prevent such combination. This is further illustrated by
the fact that the pair $\Lambda, {\Lambda_{\bf i}^{*}}(t_{{\bf
i}{\bf j}})$ may depend on the setting ${\bf i}$ in $S_1$ and the
pair $\Lambda, {\Lambda_{\bf j}^{**}}(t_{{\bf i}{\bf j}})$ on the
setting ${\bf j}$ in $S_2$ without violating Einstein locality
because the instrument parameters are allowed to depend on the
local setting.

\item[(b2)] It is very important to realize that one certainly can obtain
different averages over $u = 1,2,...,{\frac {1} {9}}Np_s$ of the
terms in Eq.(\ref{eum7}) for the nine different pairs of settings
${\bf i},{\bf j}$ because (in the best of circumstances) these
averages converge respectively to
\begin{equation}
{\int}{\int}{\int}A_{\bf i}( {\lambda}, {\lambda_{\bf
i}^{*}})B_{\bf j}( {\lambda}, {\lambda_{\bf j}^{**}})\rho({\bf i},
{\bf j}, {\lambda}, {\lambda_{\bf i}^{*}},{\lambda_{\bf
j}^{**}})d{\lambda}d{\lambda_{\bf i}^{*}}d{\lambda_{\bf
j}^{**}}\label{eum8}
\end{equation}
However, here the density $\rho$ will not only be setting
dependent but also {\it time} dependent. Thus there is no
guarantee that $\rho$ is a common factor of the nine integrands
and that the argument leading to Eq.(\ref{eum3}) will work.

\end{itemize}

It is often erroneously believed that a setting dependent density
$\rho$ automatically means Einstein non-locality. This is clearly
incorrect when time and setting dependent instrument parameters
are involved. Indeed it would be ironic if, even in the trivial
case where $\Lambda, \Lambda_{\bf i}^{*}$ and $\Lambda_{\bf
j}^{**}$ are stochastically independent, their joint density,
being now the product of the three marginal densities, would not
be allowed to depend on the settings. Of course it is well known
that the proof of the Bell inequalities does work for this special
case \cite{bellbook}. However, it is easy to give an example of
time and setting dependent instrument parameters $\Lambda_{\bf
i}^{*}(t)$ and $\Lambda_{\bf j}^{**}(t)$ in the two stations that
obey Einstein locality but are {\it not stochastically
independent} because of their common time dependence. The
corresponding density $\rho$ will depend on the settings and
Bell's proof will not work. The following construction of setting
and time dependent instrument random variables should be
instructive.

The particles emitted from the source may, in general, carry not
only information related to the spin but also other information.
As a consequence of all physical properties of the particle pair
and as a consequence of the choice ${\bf i}, {\bf j}$ of settings
by the experimenters, a pair of closely linked measurement times
$t_{{\bf i}}$ in station $S_1$ and $t_{{\bf j}}$ in station $S_2$
emerges for each measurement. Assume as before $t_{{\bf i}} =
t_{{\bf j}} = t_{{\bf i}{\bf j}}$. Different measurement times for
a given correlated pair are addressed below. The fact that we have
equal measurement times for any given correlated pair and given
pair of settings gives us the possibility to construct parameter
random variables $\Lambda_{\bf i}^{*}(t)$ and $\Lambda_{\bf
j}^{**}(t)$ that are not independent in spite of the fact that
they may be stochastically independent of the source parameter
$\Lambda$ that contains some information on the spin.
$\Lambda_{\bf i}^{*}(t)$ and $\Lambda_{\bf j}^{**}(t)$ may, for
example, exhibit a ``stroboscopic" randomness constructed on two
computers with equal clock time in the following way. Both
stations contain three stacks of files denoted by $\lambda_{\bf
a}^{*}(t)$, $\lambda_{\bf b}^{*}(t)$ and $\lambda_{\bf c}^{*}(t)$
in station $S_1$ and entirely identically arranged stacks of files
denoted by $\lambda_{\bf a}^{**}(t)$, $\lambda_{\bf b}^{**}(t)$
and $\lambda_{\bf c}^{**}(t)$ in station $S_2$. Given a particular
setting pair and measurement time $t_{{\bf i}{\bf j}}$, two actual
files are picked, one in each station. Because the stacks are
identically arranged we have for all times
\begin{equation}
\lambda_{\bf i}^{*}(t) = \lambda_{\bf i}^{**}(t) \text{   for
}{\bf i} = {\bf a}, {\bf b}, {\bf c} \label{eum9}
\end{equation}
Then pairs of settings $({\bf i}, {\bf j})$ are chosen
sequentially and at random according to an arbitrary distribution
and the measurements are performed during certain small time
periods labelled as measurement time. The possible setting and
time dependencies  i.e. the order of the files in the stacks can
be determined by arbitrary algorithms (including e.g. appropriate
elements of the history of past experiments) and determine then
the setting and time dependence of the joint probability density
of $\lambda_{\bf i}^{*}(t)$ and $\lambda_{\bf j}^{**}(t)$ for
different settings ${\bf i}, {\bf j}$. Thus, for the special case
of a time independent source parameter, the instrument parameters
may be stochastically independent of the source parameter. In
addition virtually arbitrary stochastic dependencies of instrument
and source parameters can be constructed. Next we chose the
functions $A, B = \pm1$ identical to each other and we see that
Eqs.(\ref{eum4}) and (\ref{eum5}) hold. Note that this
construction is not the most general one since Eq.(\ref{eum9}) is
only sufficient for Eqs.(\ref{eum4}) and (\ref{eum5}) to hold, but
not necessary. In summary, our extended parameter set does not
collapse onto Mermin's nor can Mermin's proof proceed in our
extended space. All his essential objections against our work that
we discussed so far are therefore without basis.

We add just two final remarks. MII tries to refute the possibility
of time and setting dependent instrument parameters also on
additional grounds by the following statement on p 146
\cite{eurom}: ``According to quantum mechanics...the statistical
character of the data...is unaffected if the two detections of a
given entangled pair are separated by arbitrary long time
intervals" and ``To maintain this in a Hess-Philipp
expansion...requires, in the absence of spookily
conspiratorial...exchange of information between the two
detectors...that the choice of...microsettings for each setting
must be the same for all times". This statement contains several
problems. First, to separate the detections of a given entangled
pair by arbitrary long times $\Delta_t$ is experimentally very
difficult. For experiments that involve optics, one would need to
try separations over a time period $\Delta_t$ covering about the
whole duration of the experiments that is necessary to accumulate
the statistical data, i.e. at least minutes. One minute multiplied
by the velocity of light corresponds to almost inter-planetary
distances for the measurements. As a consequence, such
measurements are difficult to perform and have not been performed.
Second, it is logically inconsistent to come up with conditions
that are not contained in the mathematics of Bell-type proofs and
to drag in quantum mechanics in absence of any experimental
confirmation. Third, and most importantly, our parameter space can
be adapted to derive even this additional quantum mechanical
result and without the necessity of ``spookily conspiratorial"
elements. Our argument is as follows. Any change of the
measurement in one station (wing) by a time interval $\Delta _t$
requires experimental changes that may have some causal effect on
both the source parameter $\Lambda(t)$ and the instrument
parameter $\Lambda_{\bf i}(t)$ of that wing. If, for instance, the
change occurs in the wing of station $S_1$, then all that needs to
be done is to replace $t$ by $t \pm \Delta_t$ in the time related
arguments $\{.\}$ of the function $A_{\bf a}(\{.\}, \{.\}, \{.\},
\{.\})$. A similar reasoning can be applied for the delay of
measurements in the wing of station $S_2$ or even for both wings.
The changes of the experimental configuration that need to be made
to accomplish the time shift $\Delta _t$ for a given wing in the
experiments can be taken as the causal reasons for the
transformation. This shows that Mermin's suspicion of ``spookily
conspiratorial" elements is without basis. There certainly is the
possibility that the ``conspiracy" may just be a standard
cause-effect situation.

Note finally that under Mermin's assumptions our extensive
mathematical model \cite{hpp2} can indeed be simplified. However,
these simplifications deal only with the calculations. For
example, we can replace the B-splines by more elementary
functions. The basic construction remains the same and is sketched
on the first four pages of a recent manuscript \cite{wk}. Because
we did not carry out a detailed specialization of our model as it
relates to MII, Mermin infers that our model ``must contain
errors, as has been argued directly elsewhere". In \cite{gilleur}
and \cite{gillmyr} we have refuted the papers \cite{gwzzpnas} and
\cite{myr} to which Mermin points.

Acknowledgement: the work was supported by the Office of Naval
Research N00014-98-1-0604. The authors would also like to thank
Salvador Barraza-Lopez for valuable suggestions to improve the
presentation.


\begin{thebibliography}{99}
\bibitem{cst} We denote a procedure that defies well established
mathematical protocol as countersyntaxial, indicating possible
logical errors.
\end{thebibliography}
\end{document}